\documentclass[aps,rmb,twocolumn,showpacs,superscriptaddress,showkeys,longbibliography]{revtex4-1}

\usepackage[utf8]{inputenc}
\usepackage[T1]{fontenc}

\usepackage{multirow}	
\usepackage{enumerate}	
\usepackage{Tabbing}	
\usepackage{relsize}	
\usepackage{textcomp}	
\usepackage{soul}	

\usepackage{amsmath}
\usepackage{amssymb}
\usepackage{bm}		

\usepackage{hyperref}
\usepackage{url}
\usepackage{cleveref}	

\usepackage{dcolumn}	
\usepackage{longtable}	
\usepackage{booktabs}	

\usepackage[pdftex]{graphicx}
\usepackage{epstopdf}	
\usepackage{rotating}	
\usepackage{float}	

\usepackage{listings}
\usepackage{color}
\usepackage{textcomp}

\definecolor{listinggray}{gray}{0.9}
\definecolor{lbcolor}{rgb}{0.9,0.9,0.9}

\usepackage{color}

\newcommand{\average}[1]{\ensuremath{\left\langle#1\right\rangle}}

\newcommand{\matrixel}[3]{\ensuremath{\left\langle #1 \vphantom{#2#3} \right| #2 \left| #3 \vphantom{#1#2} \right\rangle}}

\newcommand{\AN}[2]{
\ensuremath{\hat{c}^{} _{{#1}
\ifnum#2=1  \uparrow
\else
\ifnum#2=-1  \downarrow
\else
\ifnum#2=2  \sigma
\else
\ifnum#2=-2  \bar{\sigma}
\else
\ifnum#2=3  \sigma '
\fi
\fi
\fi
\fi
\fi
}}
}
\newcommand{\CR}[2]{
\ensuremath{\hat{c}^\dagger _{{#1}
\ifnum#2=1  \uparrow
\else
\ifnum#2=-1  \downarrow
\else
\ifnum#2=2  \sigma
\else
\ifnum#2=-2  \bar{\sigma}
\else
\ifnum#2=3  \sigma '
\fi
\fi
\fi
\fi
\fi
}}
}
\newcommand{\NUM}[2]{
\ensuremath{\hat{n}^{} _{{#1}
\ifnum#2=1  \uparrow
\else
\ifnum#2=-1  \downarrow
\else
\ifnum#2=2  \sigma
\else
\ifnum#2=-2  \bar{\sigma}
\else
\ifnum#2=3  \sigma '
\fi
\fi
\fi
\fi
\fi
}}
}

\newcommand{\mc}[3]{\multicolumn{#1}{#2}{#3}}

\renewcommand{\vec}[1]{\ensuremath{\mathbf{#1}}}

\definecolor{newRed}{RGB}{200,0,0}

\definecolor{newGreen}{RGB}{0,100,0}

\begin{document}

\title{Discontinuous transition of molecular-hydrogen chain to the quasi-atomic state:\\ Exact diagonalization -- ab initio approach}

\author{Andrzej P. K\k{a}dzielawa}
\email{kadzielawa@th.if.uj.edu.pl}
\affiliation{Marian Smoluchowski Institute of Physics, Jagiellonian University, 
ulica \L{}ojasiewicza 11, PL-30-348 Krak\'ow, Poland}

\author{Andrzej Biborski}
\email {andrzej.biborski@agh.edu.pl}
\affiliation {Academic Centre for Materials and Nanotechnology,
AGH University of Science and Technology,
al. Mickiewicza 30,  PL-30-059 Krak\'ow,  Poland}
 
\author{J\'{o}zef Spa\l{}ek}
\email{ufspalek@if.uj.edu.pl}
\affiliation{Marian Smoluchowski Institute of Physics, Jagiellonian University, 
ulica \L{}ojasiewicza 11, PL-30-348 Krak\'ow, Poland}
\affiliation {Academic Centre for Materials and Nanotechnology,
AGH University of Science and Technology,
al. Mickiewicza 30,  PL-30-059 Krak\'ow,  Poland}

\date{\today}

\begin{abstract}
 We obtain in a direct and rigorous manner a transition from a stable molecular hydrogen $nH_2$ single chain to the quasiatomic two-chain $2nH$ state.
 We devise an original method composed
 of an exact diagonalization in the Fock space combined with an ab initio adjustment of the single-particle wave function in the correlated state.
 In this approach the well-known problem of double-counting the interparticle interaction does not arise at all. The transition
 is strongly discontinuous, and appears even for relatively short chains possible to tackle, $n=3\div6$. The signature of the transition as a function of applied force is a discontinuous
 change of the equilibrium intramolecular distance. The corresponding change of the Hubbard ratio $U/W$ reflects the Mott--Hubbard-transition aspect of the atomization.
 Universal feature of the transition relation to the Mott criterion for the insulator--metal transition is also noted. The role of the electron correlations is
 thus shown to be of fundamental significance in this case. The long-range nature of Coulomb interactions is included.
\end{abstract}

\pacs{
31.15.A-,	
71.27.+a,	
67.80.F-,	
67.63.Gh	
}

\maketitle

\noindent
\emph{1. Motivation}
The metallization of solid hydrogen is one of the central problems in physics \cite{Wigner,Mao,Silvera,McMahon2}, as well as in astrophysics of Jupiter, Saturn, and exoplanets \cite{Baraffe}.
The building block, the $H_2$ molecule, is relatively
simple, since the electrons are in the spin-singlet state in the ground state composed mainly of $1s$ states of atoms. 
Those molecular states with the experimental value of bond length $R \approx 1.401 a_0$ \cite{Kolos} form at ambient pressure a molecular crystal with the lattice constant $a \sim 7 a_0 \gg R$ \cite{Curzon},
where $a_0$ is the Bohr radius.
Thus, to achieve atomic structure and metallicity one has to break the molecular bond, e.g., by achieving a more typical atomic solid configuration with $a \sim R$,
what amounts in practice to applying
an enormous pressure, as demonstrated repeatedly over the decades \cite{Mao,Howie}. But even then one may not achieve metallicity, as due to the relatively large
atomic separation the system may compose a Mott (or Mott-Hubbard) insulator \cite{Gebhard,Kadzielawa1}, as the original interelectronic correlation effects
may still be sufficiently strong. The fundamental aspect of this paper is to address this issue in a rigorous manner, as detailed below.

The related and probably more intriguing question related to the hydrogen metallization is the conjecture that the system may exhibit
a room-temperature superconductivity. The line of reasoning \cite{Ashcroft2, Babaev} is based on the Bardeen-Cooper-Schrieffer (BCS)
theory that as the critical temperature $T_s$ is proportional to $M^{-1/2}$, where $M$ is the ionic mass, then the hydrogen metal
should have $T_s$ at least one order of magnitude higher than that of a typical metal. However, here again the correlation
effects must be carefully taken into account as they should not be too strong to destroy the metallicity, but essential, as they
can lead by themselves also to the high-temperature superconductivity when the starting point is the atomic solid with half-filled
Mott-Hubbard metal \cite{Kaczmarczyk2}. A marriage of strong electron--lattice and correlation
effects should be accounted for at the starting stage. In this respect, here we determine rigorously the magnitude 
of the local correlation effects as described by the effective extended Hubbard model at the molecular~\textrightarrow~quasiatomic solid transition.
The values of local electron--proton coupling have been calculated elsewhere \cite{Kadzielawa2}.

Recently, the molecular--atomic hydrogen transition has been discussed by a number of methods \cite{Azadi2,Mazzola,Errea,Azadi3}. 
In this respect, the present results can be regarded as a basis for testing the approximate methods.

\begin{figure}
 \includegraphics[width=\linewidth]{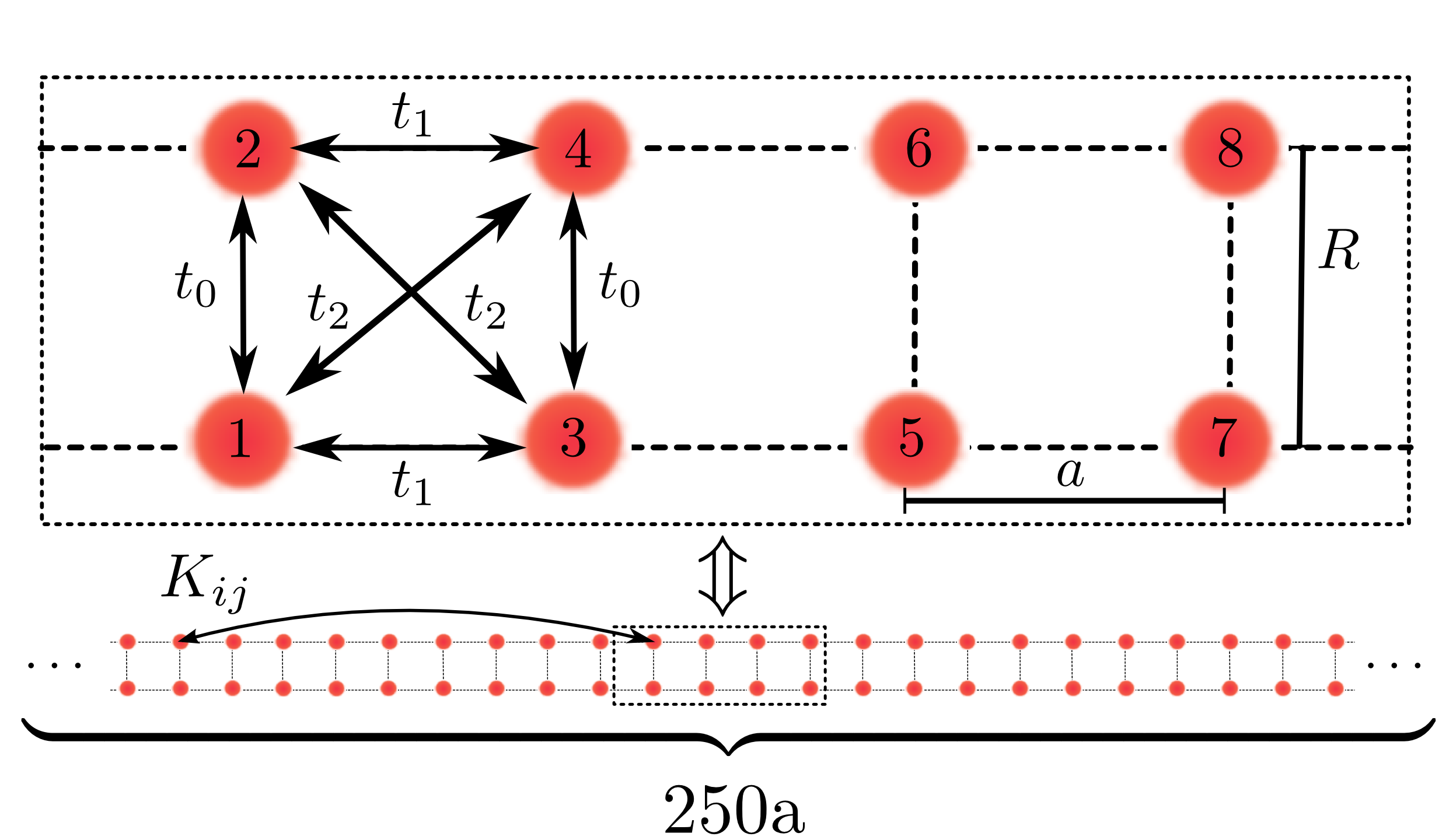}
 \caption{Schematic representation of $H_2$ molecular chain with possible nearest-neighbor and next nearest neighbor hoppings $\{t_i\}_{i=0,1,2}$ marked. The labeling of the sites $i=1,2,3,\dots$ is specified,
 as well as the bond length $R$ at intermolecular distance $a$. The radius of included intersite Coulomb interaction is equal to $250 a$ in the atomic representation. (see main text).}
 \label{fig:chain_model}
\end{figure}

\vspace{0.1cm}
\noindent
\emph{2. Model and method}
We start with the extended Hubbard model with additional term $V_{\text{ion--ion}}$ expressing ion--ion repulsion namely,
\begin{align}
 \label{eq:hamiltonian}
 \hat{\mathcal{H}} &= \sum_{i}\epsilon_{i} \NUM{i}{0} + \sum_{\sigma,i\neq j} t_{ij} \CR{i\sigma}{0} \AN{j\sigma}{0}+U\sum_{i}\NUM{i}{1} \NUM{i}{-1}\\\notag
&+ \frac{1}{2}\sum_{i\neq j} K_{ij} \NUM{i}{0} \NUM{j}{0} + V_{\text{ion--ion}},
\end{align} 
where $\epsilon_{i}$ is the single-particle energy, $t_{ij}$ are the so-called hopping integrals (all of them: $t_0$ (intramolecular) and $t_1$, and $t_2$ (intermolecular)
have been marked explicitly in Fig.~\ref{fig:chain_model}), $U$ is the on-site Coulomb repulsion, and $K_{ij}$ is the amplitude of intersite Coulomb repulsion.
The configuration of the molecules and related microscopic parameters are shown in Fig.~\ref{fig:chain_model}.
Explicitly, those parameters can be defined in the following manner
\begin{align}\notag
 \matrixel{w_i}{H_1}{w_j} &\equiv \int d^3 r \  w_i^*(\vec{r}) H_1 (\vec{r}) w_j(\vec{r}) \\
			  &\equiv \epsilon_i \delta_{ij} + t_{ij} (1- \delta_{ij},) \\\notag
 \matrixel{w_i w_j}{V_{12}}{w_i w_j} &\equiv \int d^3 r d^3 r' \  |w_i(\vec{r})|^2 V_{12} (\vec{r}-\vec{r}') |w_j(\vec{r}')|^2 \\
			  &\equiv U \delta_{ij} + K_{ij} (1- \delta_{ij}), 
\end{align}
where $H_1$ is the Hamiltonian of single electron in the medium and $V_{12}$ is the Coulomb repulsive interaction for a single
pair of them. Furthermore, the Wannier functions are defined in terms of atomic (Slater) functions, i.e., 
\begin{align}
 w_i (\vec{r}) = \sum_j \beta_{ij} \psi_j (\vec{r}),
\end{align}
where the $1s$-type Slater function is defined as $\psi_i (\vec{r}) \equiv \psi (\vec{r}~-~\vec{R}_i) = (\alpha^3/\pi)^{1/2} \exp(-\alpha |\vec{r} - \vec{R}_i|)$,
with $\alpha$ being its inverse size in the medium, here taken as a variational parameter for given intermolecular distance $a$ and determined in the correlated state
($i\equiv \vec{R}_i$ is the $i$-th proton position). Each Slater function is approximated by its expansion in the Gaussian basis
\begin{align}
 \psi_i (\vec{r}) \approx \alpha^{\frac{3}{2}} \sum_{q=1}^p B_q \left(  \frac{2 \Gamma^2_q}{\pi}\right)^{\frac{3}{4}} e^{-\alpha^2 \Gamma^2_q |\vec{r}- \vec{R}_i |^2},
\end{align}
with $p$ being the number of Gaussian functions taken into account to express accurately the functions $\{ \psi_i (\vec{r}) \}$ and $\{ B_q, \Gamma_q \}$ is the set of adjustable
parameters obtained from a separate procedure (for details see \cite{RycerzPhD}). In effect, all the parameters: $\epsilon_i$, $t_{ij}$, $U$, $K_{ij}$
can be expressed in terms of $\alpha$ and are determined in the correlated state (after the exact diagonalization in the Fock space has also been carried out simultaneously).
Strictly speaking, to obtain proper atomic limit we have to take: $\epsilon_i \rightarrow \epsilon_i^{eff} = \epsilon_i + (1/2) \sum_j (2/R_{ij} + K_{ij} )$ \cite{Biborski}.

A brief methodological note is in place here. We diagonalize exactly the clusters up to $n=5$. However, to account for the long-range nature of Coulomb interactions,
we include them in $\epsilon_i$, $t_{ij}$, $K_{ij}$, and the ion--ion interaction up to the distance $250 a$. In other words, we regard our cluser of $n H_2$ molecules
as a single block of the length $na$ immersed in the periodic system. In this manner, we utilize the periodic boundary conditions in the same sense as in the cluster
perturbation approaches, cf. e.g. Ref. \cite{Maska}.

First, we determine values of hopping integrals $t_0$, $t_1$, and $t_2$ Hubbard (intraatomic) and intersite interactions $U$ and $K_{ij}$, by determining the Wannier functions
defining them. The Wannier functions are expressed in terms of (atomic) Slater functions of adjustable size $\alpha^{-1}$, each of which is expressed in turn
via $p=9$ Gaussian functions \cite{Biborski}. In the next step, we determine the lowest eigenvalue of the Hamiltonian \eqref{eq:hamiltonian} with periodic boundary conditions,
by diagonalizing it by means of Lanczos algorithm.
By finding the ground state energy in the Fock space and subsequently, by optimizing the energy also with respect to the size $\alpha^{-1}$,
we obtain a true physical ground state energy in the correlated state for given intermolecular distance $a$. 
Such extended calculations make the results not limited to the parametrized-model considerations, as $U$, $t_{ij}$, and $K_{ij}$ are evaluated
explicitly, as is also the ground state energy, for fixed $a$. One should note that
in the procedure the bond length is also optimized ($R \rightarrow R_{eff}$). All this is carried out first for the zero-applied force to
obtain the system equilibrium configuration, i.e., the energy and the effective bond length $R_{eff}$ in the multimolecular configuration.
In the second step, we apply the external compressing force $f$ to the end molecules and determine the enthalpy minimum to trace the system evolution as a function of $a$.
The part of the procedure connected with the optimization of the single-particle wave functions has been discussed in detail elsewhere \cite{Biborski}.
The code used for the computation is also available \cite{qmtURL}.

Typical numerical procedure starts with fixing $a$ and $R$, and selecting input value for the variational parameter $\alpha$. Next, we vary the inverse wave function size $\alpha$
to find the physical ground-state energy employing, at every step, the so-called \emph{process--pool solution} for the
computationally expensive problem of obtaining the microscopic parameters. We include all the intersite interactions for the sites with a relative distance smaller than $250 a$.
The accuracy of the numerical results is set to be of the order of $10^{-4} Ry$. Impact of these parameters was examined carefully in \cite{Biborski} and achieving the values above
proved to be sufficient. The long range of the interactions in the atomic representation is expected to emulate correctly a longer-chain behavior.

\begin{figure}
 \includegraphics[width=\linewidth]{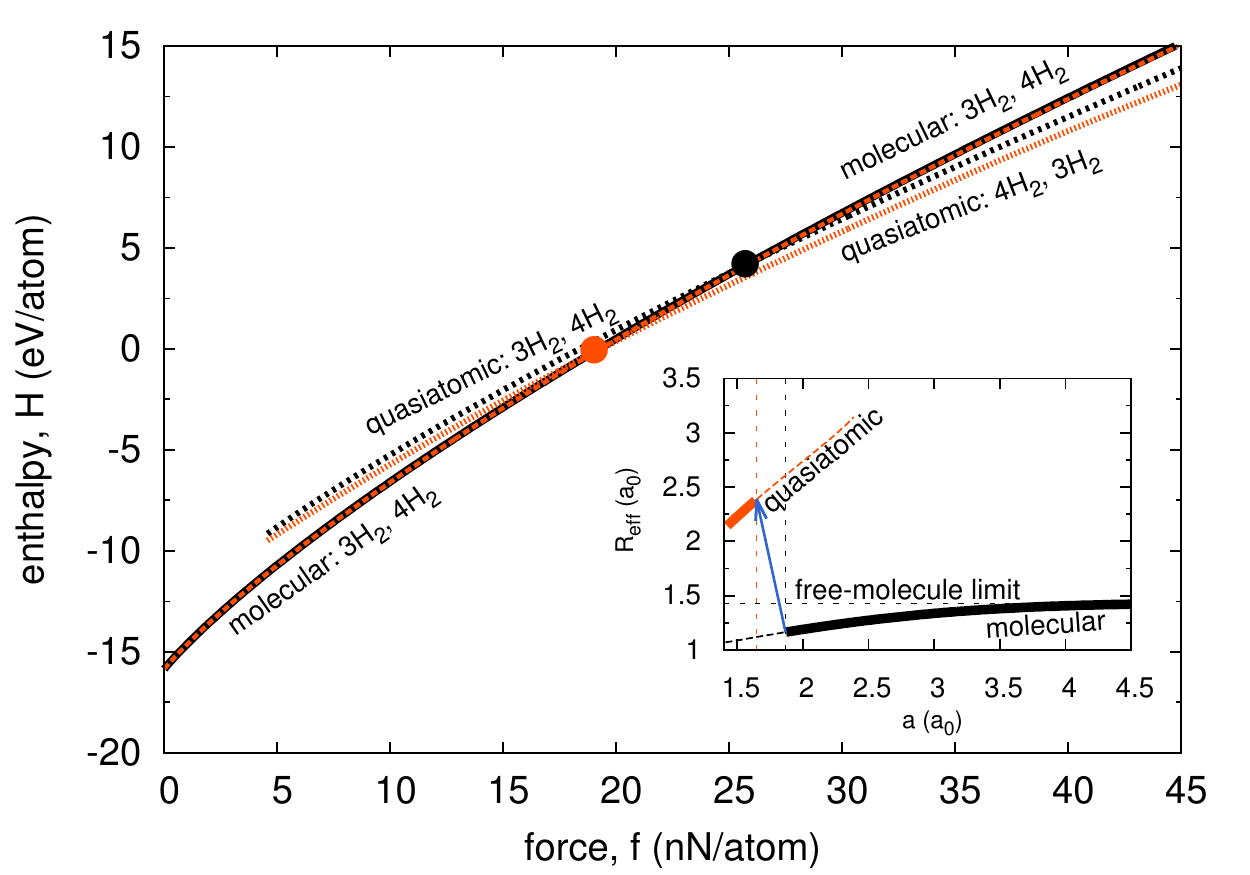}
 \caption{Phase diagram encompassing quasiatomic and molecular states for $nH_2$ chain as a function of exerted force along the chain.
	  Inset: Effective bond length $R_{eff}$ vs. the intermolecular distance $a$.}
 \label{fig:enthalpy}
\end{figure}

\vspace{0.1cm}
\noindent
\emph{3. Results}
In Fig.~\ref{fig:enthalpy} we display the system enthalpy as a function of force $f$ exerted on the molecules. 
We selected such anisotropic form of $f$, since a proper account of hydrostatic-pressure case would require a consideration of square or 3-dimensional clusters,
so the all directions are physically equivalent.
We assume a spatially homogeneous state of the system, in accordance with periodic boundary conditions. Typically, two solutions appear which we call
respectively as the molecular and the quasi-atomic, each of which is characterized below. The solid points mark the transition points
for $4H_2$ (left) and $3H_2$ (right).
\begin{table*}[t]
\caption{Microscopic parameters at the molecular~\textrightarrow~quasiatomic transition. The values are in $eV$ if not specified otherwise.}
 \label{tab:values}
\begin{tabular}{l||l|l|l|l|l|l|l|l|l|l|l}
$ $ &\mc{1}{r|}{$a$ $(a_0)$} & \mc{1}{c|}{$R$ $(a_0)$} & \mc{1}{c|}{$W$} & \mc{1}{c|}{$t_0$} & \mc{1}{c|}{$t_1$} & \mc{1}{c|}{$t_2$} & \mc{1}{c|}{$U$} & \mc{1}{c|}{$K_0$} & \mc{1}{c|}{$K_1$} & \mc{1}{c|}{$K_2$} & \mc{1}{c}{$\alpha (a_0^{-1})$} \\\hline\hline
molecular & \mc{1}{d|}{1.869} & \mc{1}{d|}{1.164} & \mc{1}{d|}{17.881} & \mc{1}{d|}{-15.487} & \mc{1}{d|}{-6.161} & \mc{1}{d|}{1.691} & \mc{1}{d|}{26.472} & \mc{1}{d|}{15.238} & \mc{1}{d|}{13.516} & \mc{1}{d|}{10.551} & \mc{1}{d}{1.194}\\\hline
quasiatomic & \mc{1}{d|}{1.647} & \mc{1}{d|}{2.386} & \mc{1}{d|}{33.124} & \mc{1}{d|}{-5.270} & \mc{1}{d|}{-8.445} & \mc{1}{d|}{0.164} & \mc{1}{d|}{26.404} & \mc{1}{d|}{10.825} & \mc{1}{d|}{14.701} & \mc{1}{d|}{8.976} & \mc{1}{d}{1.251}
\end{tabular}
\end{table*}
In the Inset we show the effective bond length $R_{eff}$ vs $a$. Note two specific features. First, the lattice
contracts in a discontinuous manner at the transition and second, the bond length jumps then from an almost single-molecule value $R_{eff} \ll a$ to the limit
$R_{eff} \sim a$ signaling their separation into a quasi-atomic configuration.
\begin{figure}
 \includegraphics[width=\linewidth]{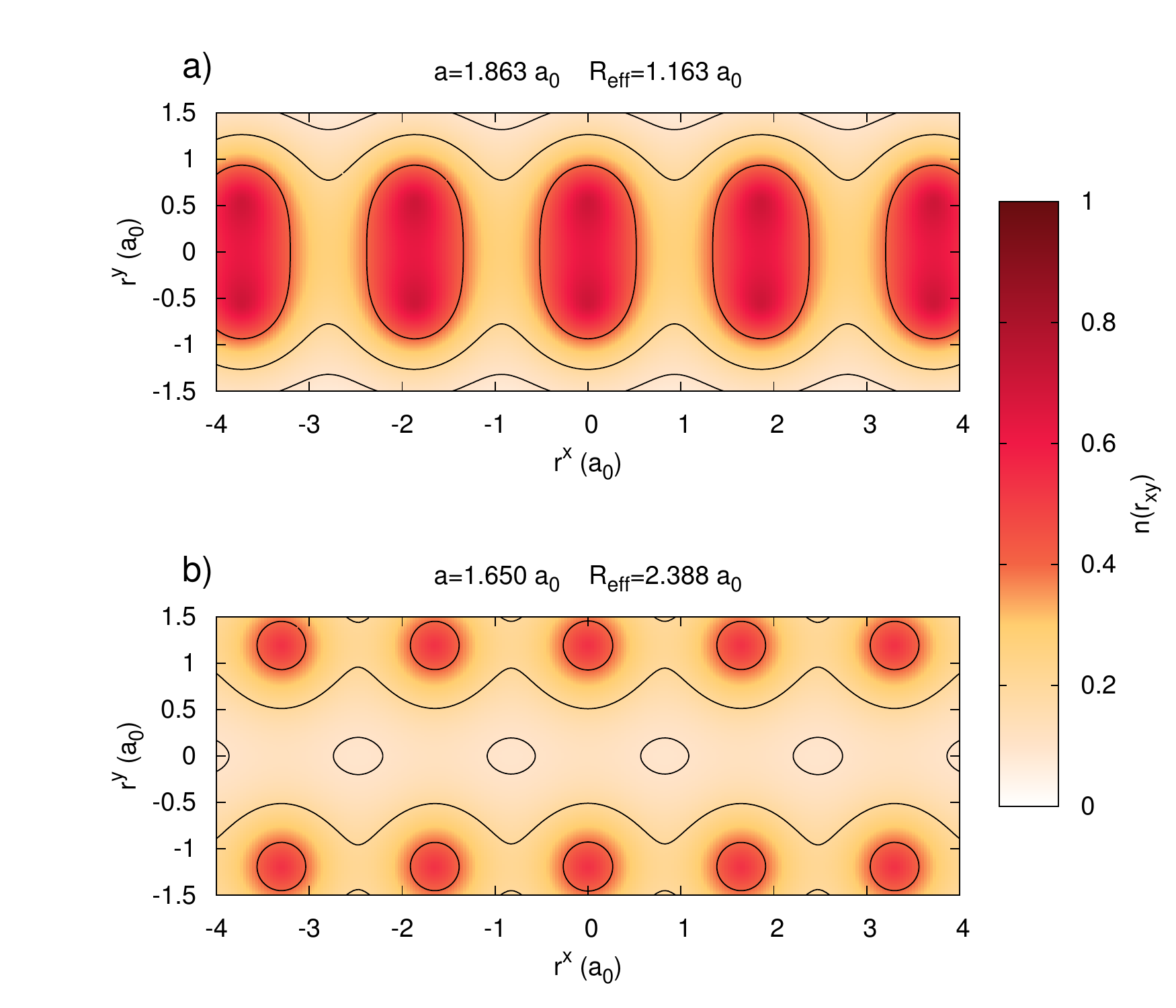}
 \caption{Electron density projected onto the plane of $H_2$ molecules near the molecular~\textrightarrow~atomic transition: a)~for the molecular state;
 b)~for the quasiatomic state. In case a) the density is very small in the space in between
 molecules (vertical line joining the atoms). The critical intermolecular distance $a$ and the effective bond length $R_{eff}$ are also specified in each of the states. The corresponding
 critical force is $f_C = 25.705 nN$ for $3H_2$ system.}
 \label{fig:maps}
\end{figure}
The last feature of the transition is explicitly illustrated in Fig.~\ref{fig:maps}, where the projected electron density onto the plane composed of molecules
is shown both on the molecular (top) and the quasi-atomic (bottom) sides of transition. The parameters $a$ and $R_{eff}$ are listed also in each case.

\begin{figure}
 \includegraphics[width=\linewidth]{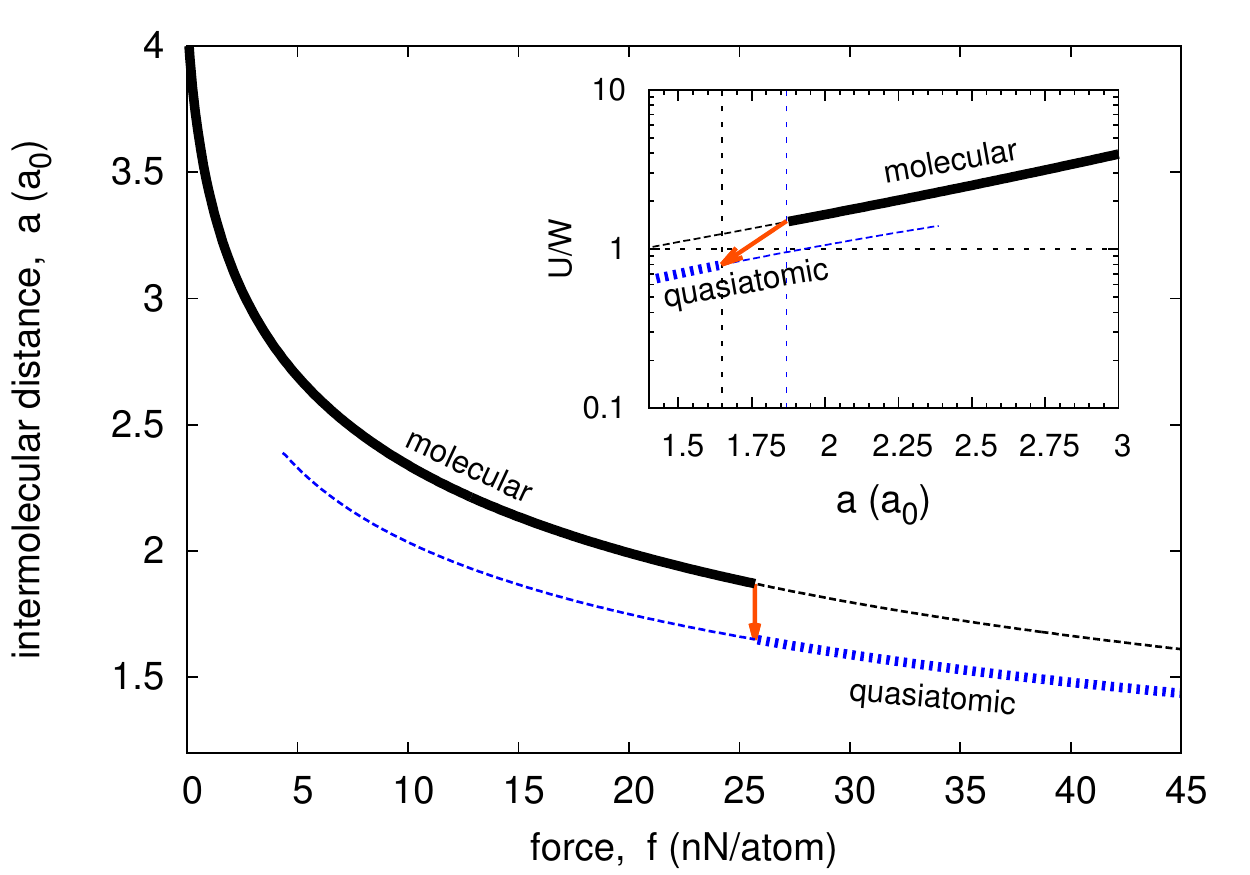}
 \caption{Equilibrium intermolecular distance $a$ versus exerted force $f$ for the case of $3H_2$ system. The transition is of the first order as marked.
 Inset: $U/W$ ratio vs. $a$ showing that in the quasiatomic state $U/W \sim 0.8$ signaling the onset of a correlated metallic state.}
 \label{fig:sizes}
\end{figure}

In Fig.~\ref{fig:sizes} the interdependence of the intermolecular distance versus force is provided. As shown also in the inset to Fig.~\ref{fig:enthalpy}, this figure
demonstrates the first-order transition, since the cell volume $a$ changes discontinuously at this mol.~\textrightarrow~quasiat.
transition from $1.869 a_0$ to $1.647 a_0$. In the inset to Fig.~\ref{fig:sizes} we provide the ratio of the Hubbard interaction $U$ to the bandwidth defined 
as $W = -4 t_1 + \left| t_0 + 2 t_2 \right| -  \left| t_0 - 2 t_2 \right|$, as a function of $f$.
Note the fundamental characteristic: this ratio jumps from the value $U/W \sim 1.49$ to $0.8$. So, the system is indeed strongly correlated at the transition
with the effective gap near the transition from the molecular side estimated as $\epsilon_g/W =  U/W - 1 \approx 0.49$.
Furthermore, the value $U/W \sim 1$ is the typical value for the Mott-Hubbard transition, weakly dependent on the system structure \cite{Datta}. It is also
interesting to note that here the Mott-Hubbard transition takes place from a non-magnetic (spin-singlet) molecular configuration to a correlated atomic solid,
with strong suggestion for metallicity ($U/W < 1$) of such system with one electron per atom.

\begin{figure}
 \includegraphics[width=\linewidth]{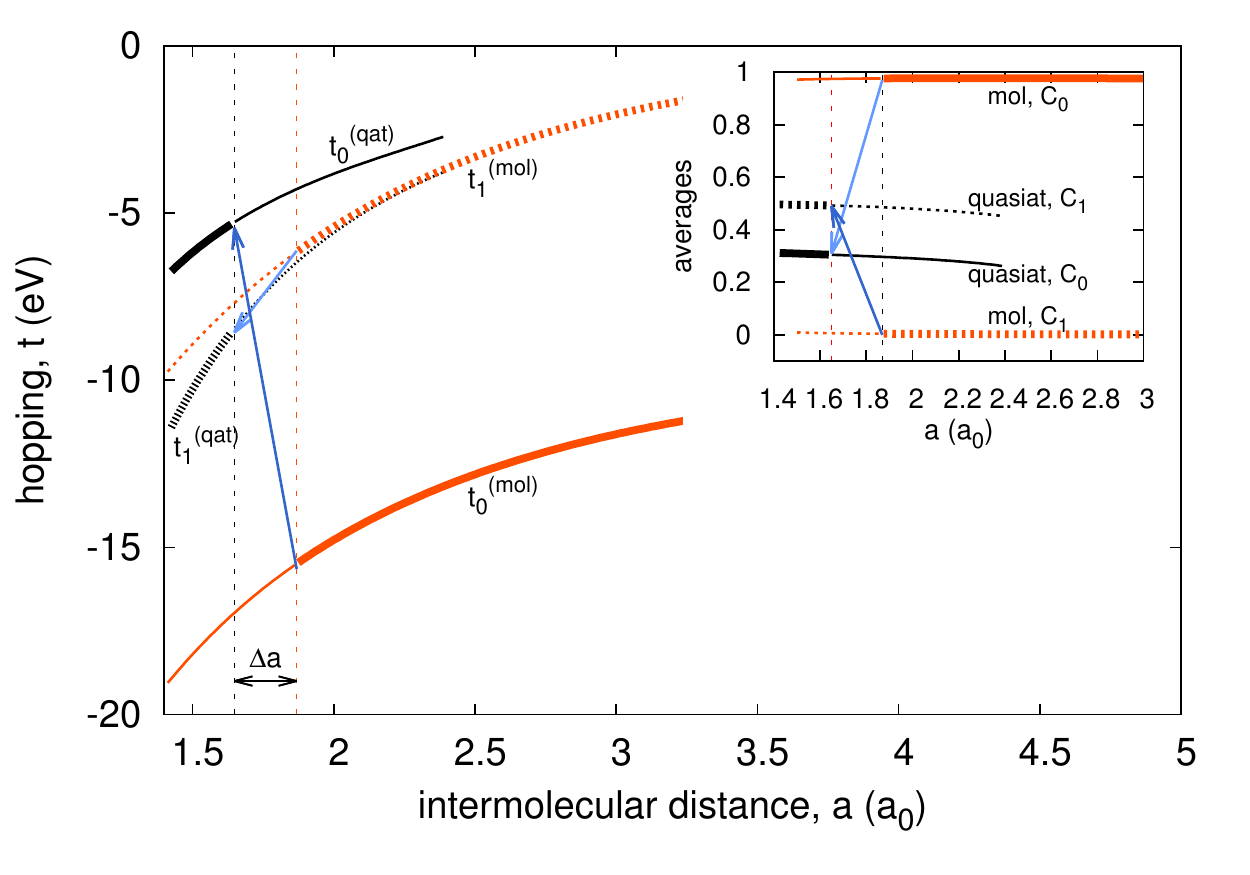}
 \caption{Evolution of the intra- and inter-molecular hopping parameters, $t_0$ and $t_1$ respectively, versus intermolecular distance $a$. Note the jump of $a$
 at the transition molecular~\textrightarrow~quasiatomic configurations as marked by the arrows. Inset: hopping probabilities: intra- and inter-molecular, $C_0$ and $C_1$, respectively.
 For discussion see main text.}
 \label{fig:hoppings}
\end{figure}
In Fig.~\ref{fig:hoppings} we present $a$ dependence of the intramolecular ($t_0$) and intermolecular ($t_1$) hopping parameters. At the transition the parameters
$t_0$ and $t_1$ roughly equalize showing again that the solid at small $a$ is of quasi-atomic nature. In the inset we present the intramolecular ($C_0 \equiv \average{\CR{1}{2}\AN{2}{2}}$)
and intermolecular ($C_1 \equiv \average{\CR{1}{2}\AN{3}{2}}$) hopping probabilities. In the molecular state $C_0 \approx 1$ and $C_1 \approx 0$ 
(a strong molecular bond is formed), whereas in the quasiatomic state $C_0 \sim C_1$. These two quantities illustrate thus in a direct manner the nature
of the states and in particular, the bonding in the quasiatomic solid state. For the sake of completeness, in Table~\ref{tab:values} we list all the principal microscopic
parameters in the correlated state for intermolecular distances at transition. The values of $K_{ij}$ converge towards the asymptotic value $K_{ij} \rightarrow 0$ slowly.
Such behavior is in accordance with the view \cite{Hubbard2} that the long-range part of the Coulomb interaction is not screened efficiently
in low-dimensional systems. This is the reason why we have taken their long-range character.
Parenthetically, note that we have neglected the direct exchange interaction and the so-called correlated-hopping terms \cite{Spalek1}, as they are
of much smaller amplitude at these distances. Note also that the Mott criterion at the transition \cite{Mott,Spalek4} takes in the present situation
the form $a_B n_C ^{1/d} \equiv (\alpha a )^{-1} \approx 0.45$ at the transition, not too far from the value $0.25$ for $3$-dimensional ($d=3$) systems.

\vspace{0.1cm}
\noindent
\emph{4. Outlook}
We have described in a rigorous manner the behavior of electrons in the short $H_2$ chains connected via the Coulomb interaction with
the further molecules, as well as utilizing the periodic boundary conditions, emulating together
an extended-chain behavior. Amazingly, the results obtained meet some of the features one can expect for a three-dimensional hydrogen molecular \textrightarrow~quasiatomic 
transformation. Obviously, no detailed realistic phase diagram can be obtained, as both the structural details and the effect of zero-point motion
are still missing. However, a unique connection to the Mott-Hubbard transition is established. A direct proof of metallicity of the quasiatomic state would involve
the determination of optical conductivity \cite{Rozenberg}. The model formulated here may form a basis for quantitative approach to solid hydrogen
with a proper account of electronic correlations. In such approach a three-dimensional realistic model must be constructed. We think that tackling
of this problem is possible on a small scale within our \textbf{E}xact \textbf{D}iagonalization \textbf{Ab I}nitio (EDABI) approach employed here and to other problems \cite{Spalek2}
involving electronic correlations as an essential aspect of their physical properties. 
For $3d$ system, approximate second-quantization diagonalization schemes such as the statistically-consistent Gutzwiller approximation may be used
with concomitant single-particle wave-function optimization \cite{Spalek4}. These methods can be tested in the present exact limit.

One should also note that the principal assumption made at this point
is that the protons form a lattice even in the atomic phase, not a proton-electron quantum-liquid plasma \cite{Mazzola,Morales,Bonev,Nellis}. Such problem
is possible to tackle within the present model by assessing the optimal ground-state energy for a random choice of the proton positions. But first,
an accurate estimate of the zero-point motion amplitude and the corresponding energy \cite{Kadzielawa2} must be carried out for an extended system with correlations \cite{Spalek3}.

Computer animation of the molecular~\textrightarrow~quasiatomic transition is attached as Supplemental material \cite{supplementary_material}.

\vspace{0.1cm}
\noindent
\emph{5. Methodological remark}
The correlated ladder systems modeled by the (extended) Hubbard model have been studied extensively in the past (cf. e.g. \cite{Tsuchiizu}).
Those studies differ from the problem discussed here. We start with the physical $H_2$ spin-singlet-insulator chain. Under pressure, the spin-singlet
states transform into a quasi-atomic correlated state. Most importantly, we treat the correlations within the cluster of $n$ molecules rigorously and at each step
determine the ladder rung size (bond length $R$) and the equilibrium lattice constant $a$. Such analysis has not been carried out so far
and provides also the values of the microscopic parameters.

\vspace{0.1cm}
\noindent
\emph{6. Acknowledgments}
We are grateful to Drs. Adam \mbox{Rycerz} and Marcello Acquarone, and to Prof. Maciej M. Maśka for helpful discussions. The work was supported by the National Science Centre (NCN),
Grant No. DEC-2012/04/A/ST3/00342.

\newpage

\bibliography{bibliography} 

\end{document}